\documentclass[twocolumn, switch]{article} %

\usepackage{preprint}

\usepackage{amsmath, amsthm, amssymb, amsfonts}

\usepackage[numbers,square]{natbib}
\usepackage[utf8]{inputenc}	%
\usepackage[T1]{fontenc}	%
\usepackage{xcolor}		%
\usepackage[colorlinks = true,
            linkcolor = purple,
            urlcolor  = blue,
            citecolor = cyan,
            anchorcolor = black]{hyperref}	%
\usepackage{booktabs} 		%
\usepackage{nicefrac}		%
\usepackage{microtype}		%
\usepackage{lineno}		%
\usepackage{float}			%

\usepackage{lipsum}		%

\usepackage{listings} %

\usepackage{newfloat}
\DeclareFloatingEnvironment[name={Supplementary Figure}]{suppfigure}
\usepackage{sidecap}
\sidecaptionvpos{figure}{c}

\usepackage{titlesec}
\titlespacing\section{0pt}{12pt plus 3pt minus 3pt}{1pt plus 1pt minus 1pt}
\titlespacing\subsection{0pt}{10pt plus 3pt minus 3pt}{1pt plus 1pt minus 1pt}
\titlespacing\subsubsection{0pt}{8pt plus 3pt minus 3pt}{1pt plus 1pt minus 1pt}

\title{DICODerma: A practical approach for metadata management of images in dermatology }

\usepackage{xwatermark}
\newwatermark[firstpage,color=gray!90,angle=0,scale=0.28, xpos=0in,ypos=-5in]{*correspondence: \texttt{eapenbp@mcmaster.ca}}

\usepackage{authblk}

\author[1]{Bell Raj Eapen}
\author[2]{Feroze Kaliyadan}
\author[3]{Ashique Karalikkattil T}

\affil[1]{McMaster University, Hamilton, Ontario, L8S 4L8, Canada}
\affil[2]{Faculty of Dermatology, King Faisal University, Saudi Arabia}
\affil[3]{Amanza Skin Clinic, Perinthalmanna, Kerala, India}

\begin{document}

\twocolumn[ %
  \begin{@twocolumnfalse} %

\maketitle

\begin{abstract}

\IfFileExists{paperaj/abstract.tex}{Clinical images are vital for diagnosing and monitoring skin diseases,
and their importance has increased with the growing popularity of
machine learning. Lack of standards has stifled innovation in
dermatological imaging, unlike other image-intensive specialties such as
radiology. We investigate the meta-requirements for utilizing the
popular DICOM standard for metadata management of images in dermatology.
We propose practical design solutions and provide open-source tools to
integrate dermatologists' workflow with enterprise imaging systems.
Using the tool, dermatologists can tag, search, organize and convert
clinical images to the DICOM format. We believe that our less disruptive
approach will improve the adoption of standards in the specialty.
}{abstract here}

\end{abstract}
\vspace{0.35cm}

  \end{@twocolumnfalse} %
] %

\IfFileExists{inclusions.tex}{
\section{Introduction}

Dermatology being a visual specialty, dermatologists rely on images for documenting and evaluating patient outcomes. However, unlike
radiology that relies widely on accepted standards for imaging, dermatologists lack standardized methods for acquisition, transfer and
archival of clinical images \citep{Eapen2020i}. The lack of standardisation has been a major drawback when it comes to
large-scale imaging and documentation in dermatology. With machine learning (ML) gaining momentum and popularity in the recent times, the
need for standardised digital imaging has also increased. Many of these emerging ML methods need efficient and effective management of
images for training, testing and validating models.

The lack of a well-established standard has an impact on patient privacy as well \citep{Eapen2020j}. Dermatologists do not
have standards-based solutions such as the Picture Archival and Retrieval System (PACS) to rely on for sharing images among them and peers.
Hence, they are often compelled to resort to less secure methods such as email and social media platforms. Most dermatologists rely on their
own personal methods for image archival. Hence, they find it difficult to compile or retrieve images belonging to a specific category
(example: images of mucosal lesions) for discussions, presentations or any academic activity, a task which is very easily done by their
radiology colleagues.

Digital Imaging and Communications in Medicine (DICOM) is a widely accepted and comprehensive standard for image acquisition, transmission
and storage in radiology and related specialties. Most devices for image acquisition and display support the DICOM standard. Much work has
been done to port the DICOM standard to dermatology, but the efforts so far have been largely unsuccessful
\citep{Caffery2018}. The consistent display of an image is less critical in dermatology for diagnosis and the imaging needs
are (or traditionally were) less intensive compared to radiology. This led to the resistance in adopting DICOM -\/- a comprehensive and
complex standard for image management. Unlike standard consumer image file formats such as JPEG and BMP, DICOM supports the storage of
clinical metadata such as the patient demographics along with the image. Traditionally dermatologists rely on auxiliary systems such as the
electronic medical records (EMRs) for the clinical metadata.

DICODerma is a tool and a preliminary standard to reconcile the best of both worlds --- the simplicity of consumer image tools and the
DICOM and PACS based enterprise imaging infrastructure. DICODerma can encode some of the relevant DICOM tags in the EXIF (Exchangeable Image
File Format) header space of ordinary digital images. Using DICODerma we built a plugin for the popular open-source image viewer for
healthcare --- ImageJ --- to manage these metadata in digital images in dermatology. ImageJ has been used previously in
dermatological applications such as constructing three‐dimensional images from optical coherence tomography \citep{Cao2015}
and quantifying allergic and irritant patch test reactions \citep{Ohshima2014}. Using our plugin called DIT4IJ, metadata can
be added to any digital image, search images based on the metadata and convert ordinary digital images to the DICOM format. DIT4IJ stands
for Dermatology Image Tagger for ImageJ.

DIR4IJ allows dermatologists to use the existing tools that they are familiar with, and at the same time leverage some of the advantages of
an enterprise imaging infrastructure such as greater patient privacy, patient safety, and better compliance with legislative requirements
for image retention.

The rest of the article is structured as follows. First, we briefly describe the DICOM specifications and the associated terminologies and
how they pertain to dermatology. Then we systematically explore the meta-requirements for extending the DICOM standard to dermatology based
on our personal experience. Next, we describe our meta-design --- a java library for storing and retrieving patient metadata as EXIF
tags called DICODerma. Then we describe how we used DICODerma to build an ImageJ plugin for dermatologists (DIT4IJ) to tag and organize
images and to convert them to the DICOM format. Finally, we discuss some of the advantages and limitations of our approach.

\section{The DICOM Standard}

DICOM is one of the most widely used standards in healthcare defining formats for images and structured data, workflow management and
network protocols \citep{Bidgood1996}. The National Electrical Manufacturers Association (NEMA) foresees the administration
of the standard but has no license requirement for use. Some of the common terms associated with DICOM are the service object pair (SOP) and
the image object definition (IOD). Though IODs are generic classes, most IODs represent individual real-world entities such as X-rays and
MRI along with the associated metadata. The combination of an IOD with a service such as storage, print or query, is the SOP.

The various metadata associated with the images includes patient demographics, series (a group of closely related images), study (all series
associated with one procedure) and the acquired binary image data. The metadata has a numerical key called the tag, data type called the
value representation (VR) and the value multiplicity (VM) count. The metadata is organized into logical groups such as the patient module.
The list of these specifications that a product supports is called the conformance statement. In short, DICOM specifies storage for storing,
processing, transmitting and displaying imaging data. The DICOM header is seen in  Figure \ref{Figure_1}.

\section{Imaging standards in Dermatology}

Imaging standards have a crucial role in the clinical image management in dermatology owing to its highly visual nature. Dermatologists use
different types of images ranging from dermoscopy to total-body maps. Sophisticated methods such as reflectance confocal microscopy are also
becoming increasingly popular. In this article, we give emphasis to the common digital photographs, but some of the discussions may apply to
other modalities as well.

Image metadata is important in dermatology as in other domains. The useful metadata includes demographic details, clinical findings, device
settings and image characteristics. Accurate rendering of images and acquisition context is important in dermatology as well
\citep{Caffery2018}. Dermatology has a distinct ontology that is used for an accurate textual description of lesions. The
metadata standards should support the domain-specific ontology of dermatology and support the emerging modalities.

Though dermatology is highly visual, dermatologists do not completely rely on the captured images for diagnostic, prognostic and therapeutic
decision making, and as such accuracy of colour and resolution is not very crucial. Images are mainly used for documentation, but with the
increasing popularity of teledermatology, parameters like resolution and color accuracy have also become more important. Dermatologists,
especially those working in the community and those in limited resource settings, rely on consumer devices such as digital cameras and
smartphones for image capture and documentation. Image capture mostly happens during a face-to-face consultation and routine physical
examinations. Hence, though the DICOM standard can be used as it is in dermatology, its overall adoption by vendors as well as practitioners
has not been very encouraging as of now. The lack of adoption is mostly due to the large overhead required for the implementation and
adoption of a comprehensive standard such as DICOM.

The workgroup 19 (WG19) of the DICOM consortium has explored ways in which DICOM can be extended to dermatological applications though the
group did not propose a complete final standard \citep{Caffery2018}. The existing IODs such as the Visible Light (VL) and the
Standard Capture (SC) can be used for dermatological applications with little modifications. Device and acquisition-related metadata are
captured by consumer-devices and encoded in the EXIF header supported by many digital image storage formats. There is some overlap between
EXIF and DICOM header tags.

\subsection{The Machine Learning revolution}

The growing popularity of machine learning (ML) and artificial intelligence (AI) applications in dermatology has brought new requirements
for image management \citep{Eapen2020i}. The need for standardized images, labelled with appropriate metadata, is an enabler
for AI applications. The digital revolution encourages sharing of images with peers and experts from other disciplines for opinion and as
such being part of the wider institutional image management infrastructure such as the picture archiving and communication system (PACS).
Adoption of Electronic Health Record (EHR) systems made it necessary to have a complete digital longitudinal patient record that includes
clinical images captured during a dermatology encounter. The need for adopting enterprise-imaging standards is becoming increasingly
important in dermatology.

\section{Our Approach}

Guided by the design science research methodology \citep{Hevner2004}, we systematically investigated the solution space for
the problem of standardizing the digital image workflow for dermatology. Our aim was to find generalizable design knowledge that can guide
system designers and policymakers. Though specific requirements vary among different user groups of an information system, they follow
generic laws called meta-requirements \citep{Kaiya2018}. We identified some of the meta-requirements as below:

\begin{enumerate}
\def\labelenumi{\arabic{enumi}.}
\item
  The existing DICOM standard should be leveraged as much as possible so that existing solutions such as PACS can be directly used in
  dermatology.
\item
  The users should be able to enter the DICOM ecosystem without adopting the entire standard, ideally using simple tools that are already in
  use.
\item
  The solution should be usable even with no vendor adoption, but vendors who adopt the standard should have an incentive to do so.
\item
  The solution should support improved patient privacy.
\item
  Search Engine Optimisation {[}SEO{]}: Search engines and social media platforms have an increasingly important role in knowledge
  dissemination in a privacy-preserving manner. Potential solutions should address the needs of these platforms
  \citep{Sharifzadeh2020}.
\item
  The standard should support emerging techniques such as machine learning and artificial intelligence.
\item
  The meta-design should be sufficiently abstract so that it can be easily implemented by vendors and users to support new needs.
\item
  The standard should be simple and easy to adopt and adapt to, leveraging existing tools.
\end{enumerate}

\section{Design}

As potential users of DICODerma, we adopt a meta-design approach to translate the generalizable meta-requirements as described above into a
prototype that can be extended. We created two software artifacts (meta-design) in the solution space that aligns with the above
meta-requirements. One is a java library called DICODerma, to encode some of the important DICOM tags as EXIF tags. The other is a plugin
called DIT4IJ for the popular open-source biomedical image management software --- ImageJ. Both are open-source available from the
GitHub repository \citep{Eapen2021}. Before we describe our meta-design in detail, we will briefly introduce the EXIF
standard and the ImageJ platforms that form the building blocks for our meta-design.

\begin{figure*}
\centering
\includegraphics[width=5.31643in,height=3.98732in]{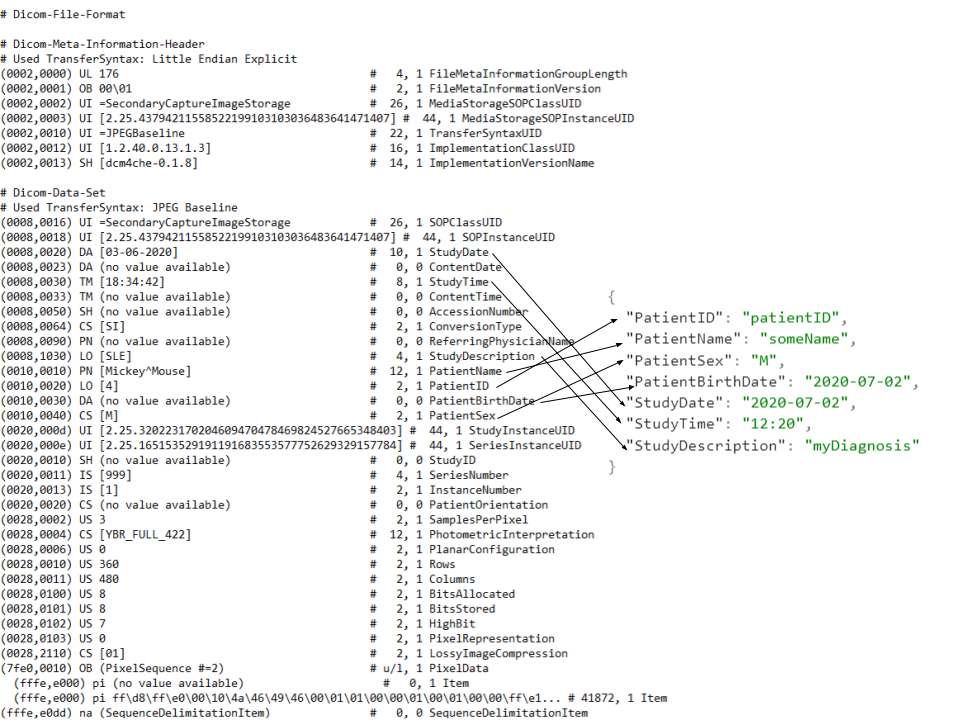}
\caption{Mapping of DICOM tags to JSON for inclusion in the UserComment EXIF tag}
\label{Figure_1} 
\end{figure*}

\subsection{EXIF Tags}

EXIF tags (hereafter EXIF) are metadata tags added by consumer devices such as digital cameras to digital images captured by these devices
(this includes images captured on smartphones too). EXIF captures a variety of details ranging from date and time information to camera
settings such as aperture and shutter speed, and GPS coordinates for the location of capture. EXIF is part of the TIFF specification and can
be found in image file types such as JPG and PNG in addition to TIFF. The GIF format does not support EXIF. Some tags such as the EXIF
version are mandatory while most tags are optional such as the user comment tag. EXIF is a consumer specification and does not support any
of the clinical tags in the DICOM header. However, some of the EXIF tags overlap with headers in the DICOM IODs. We adopt a design approach
that leverages the EXIF for clinical tags.

\subsection{ImageJ}

ImageJ is an image analysis program developed by the National Institute of Health (NIH), widely used for biomedical image analysis
\citep{Schneider2012}. ImageJ is an open-source JAVA-based software with an extensible plug-in architecture. The first
version which was released 25 years back was rewritten as ImageJ2 with additional functionalities. ImageJ2 and Fiji (ImageJ bundled with a
range of plugins that facilitate scientific image analysis) are widely used for biomedical image management
\citep{Schindelin2012}.

DICOM SC IOD is for images that are converted from a non-DICOM format such as JPEG and PNG. It is a modality independent DICOM format with
no constraints on the pixel data format. Though the initial specification was confined to single-frame images, it has been expanded to
include multi-frame images. As SC IOD is modality independent PACS will not assign any modality \citep{Pianykh2012}.

We mapped common demographic and study-related tags from the DICOM SC IOD to a JSON structure as shown in Figure 1. The DICODerma Java
library (hereafter DICODerma) facilitates writing the JSON, represented as a string, to the `UserComment' section of EXIF. DICODerma can
read and parse the JSON string from EXIF. This enables mapping useful DICOM tags to EXIF enabling the inclusion of patient metadata in
consumer image files. DICODerma uses popular and open-source dcm4che java library \citep{Warnock2007} for writing DICOM (dcm)
files from JPEG file format, a popular format supported by most capture devices and image editing software. These converted DICOM files can
be used in any system that supports these standards.

\subsection{DIT4IJ}

ImageJ has several plugins that can display, edit, save and process digital images in various formats including DICOM. Owing to the
extensible, plugin architecture of ImageJ, advanced uses not natively supported by ImageJ can be added. The modules are typically written in
Java and can be installed from the ImageJ user interface or manually copied to the plugins folder in the ImageJ folder structure. The
additional functions introduced by the plugins can be easily integrated into the ImageJ graphical user interface (GUI). The plugins,
depending on their type and functions, implement certain abstract base classes in the ImageJ core and provide implementations for methods
such as \emph{run} and \emph{setup}.

\begin{figure*}
\centering
\includegraphics[width=5.23958in,height=4.47348in]{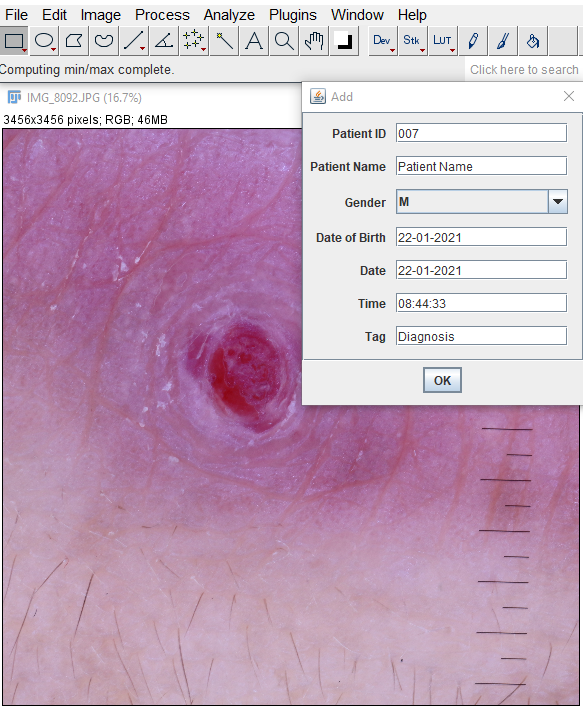}
\caption{The DIT4IJ interface for adding tags to an image}
\label{Figure_2} 
\end{figure*}

DIT4IJ is an ImageJ plugin that adds the following four functions as submenus in the ImageJ. The `add tags' function receives the tags
-\/-\/- patient id, patient name, gender, date time and diagnosis --- from the user and converts them to a JSON string and writes the
string to the `User Comment' EXIF tag of an image. The `StudyDescription' tag is used to capture the diagnosis ( Figure \ref{Figure_1}). The ImageJ
provides the interface for inputting these tags ( Figure \ref{Figure_2}). DIT4IJ can display these tags for any image and provides an interface to search
for these tags in a folder structure. For example, it can open all images of a particular diagnosis such as lichen planus by searching in
any specified file folder in the computer, including all subfolders in the search. The consumer file formats such as JPEG can be converted
into DICOM and saved anywhere in the system. This converted DICOM (dcm) file can be used with any DICOM aware application. See the attached
video file to see the usage demonstration.

\subsection{Advantages}

We address the common limitation in the existing consumer image formats --- the lack of support for patient metadata. This need is
addressed without affecting the images by the use of EXIF. The clinicians can still continue to use their imaging tools for capture,
processing and visualization of images. Some of the visualization tools support viewing the EXIF metatags including UserComment, though the
JSON formatted string is not meant for direct visualization.

We introduce ImageJ, a popular biomedical imaging software to the dermatology community. ImageJ is currently not a popular image viewer for
clinical dermatology though it has use in dermatopathology. Some of the image manipulation algorithms for clinical and cosmetic dermatology
can be easily built using the modular and extensible ImageJ framework. Some such commercial products are available
\citep{Xu2016}. We believe that the functions introduced by DIT4IJ will make ImageJ, a useful tool in dermatologists'
armamentarium and democratize imaging workflows.

The adoption of the DICOM standard in dermatology depends a lot on the vendor support and the incorporation into commercial software
products. The open-source dicoderma library could facilitate the adoption of these standards by the software vendors. Dermatologists
increasingly use smartphones as a handy image capture device. DICODerma can be used in smartphone apps to provide image tagging capability.

The inclusion of patient metadata in consumer file formats may violate patient privacy if these images are inadvertently shared. The
metadata can be anonymized using the same techniques used for anonymizing DICOM resources. With wider adoption of this standard, patient
privacy may paradoxically improve as EXIF can be easily checked for the presence of dicoderma tags. The sharing platform can reject or block
these images if these tags are present. For example, social media platforms can automatically reject any uploaded image if that image has
the dicoderma tags in them.

The dicoderma tags will facilitate machine learning. One of the challenges with machine learning in dermatology is the lack of availability
of labelled images in a privacy-preserving manner. Currently, labels associated with images should be supplied as a separate file with
unique identifiers. This is not ideal for collaboration and sharing of resources between teams. Images with dicoderma tags can be processed
and tags extracted without the need for maintaining an associated metadata file.

\subsection{Limitations}

DICODerma can only handle JPEG images with the traditional EXIF structure. Sources that generate other file types such as PNG and GIF cannot
be used with DICODerma. DICODerma uses the dcm4che library \citep{Warnock2007} to convert JPEG images to compressed DICOM
files. All DICOM readers do not yet support compressed DICOM files. The chance of inadvertently sharing sensitive patient information is a
challenge in this method though encryption of EXIF is a solution, again at the cost of increasing the the complexity
\citep{Talirongan2018}. DICODerm needs further development to support other modalities such as dermoscopy and optical
coherence tomography.

The SC IOD is a general-purpose IOD for use with any digital image. As the SC IOD is not associated with any modality, some PACS systems may
not handle them well. SC IOD lacks the meta-data model to cater to dermatologists' unique needs such as patient positioning and lighting.
However, unlike other specialties that need specialty-specific metadata model, the dermatological community's needs may be minimal. The
machine-learning algorithms may be less tolerant of variability in colour and lighting than human observers, and these requirements may
change in the future \citep{Badano2015}. We believe that our approach will introduce dermatologists to the many advantages of
standardization and ignite interest in developing a specialty-specific IOD in the future.

\section{Discussion}

The standardization requirements for dermatological images are beyond the handling of patient metadata. The proposed method of using EXIF
and interconversion with DICOM header fields are easily extensible to capture other relevant metadata. Mainstream
\citep{Amri2014} and specialized search engines \citep{Tschandl2019} are becoming increasingly accurate and
useful for dermatologists and residents. DICODerma method can improve the accuracy further because of the availability of standard metadata.

Teledermatology is becoming increasingly important because of the scarcity of dermatologists, especially in resource-poor areas. The
exchange of good quaity clinical images between patients and dermatologists is vital in teledermatology
\citep{Chuchvara2020}. The discussions related to skin findings in pandemics such as COVID-19 is crucial for screening.
DICODerma may improve the efficient use of images for these purposes \citep{Gorrepati2020}.

Smartphone based image acquisition is the new normal in dermatology, with dermoscopic addons becoming available for handheld devices
\citep{Tognetti2020}. Standardizing image capture from handheld devices along with relevant metadata, is the need of the
hour. Vendors can incorporate simple solutions using DICODerma in apps that dermatologists routinely use \citep{Jacob2020}.

We propose a simple method and tool for managing imaging metadata in dermatological images. Our method is suitable for an encounter-based
workflow, commonly seen in dermatology. In an encounter-based workflow, the imaging forms part of other clinical documentation, unlike in an
order-based workflow where the image-acquisition may be the primary purpose of the visit \citep{Cram2016}. The possibility of
integrating with the enterprise imaging systems with minimal change to the traditional and straightforward imaging methods that
dermatologists are used to might lead to the development of more elaborate standards.

}{ The chapters here. }

\normalsize
\bibliography{references}

\end{document}